\documentclass[prd,aps,showpacs,nofootinbib,nobibnotes,superscriptaddress]{revtex4}
\usepackage{epsfig}
\usepackage{amssymb}
\usepackage{graphics}
\usepackage{bm}

\begin{document}

\title{Faraday Rotation Limits on a Primordial Magnetic Field from WMAP Five-Year Data}

\date{\today}

\author{Tina Kahniashvili}
\email{tinatin@phys.ksu.edu} \affiliation{Department of Physics and McWilliams Center for Cosmology,
Carnegie Mellon University, Pittsburgh, PA 15213, USA}
\affiliation{Department of Physics,
Laurentian University, Ramsey Lake Road, Sudbury, ON P3E 2C6,
Canada} \affiliation{National Abastumani Astrophysical
Observatory, Ilia Chavchavadze State University, 2A Kazbegi Ave,
Tbilisi, GE-0160, Georgia}

\author{Yurii Maravin}
\email{maravin@phys.ksu.edu} \affiliation{Department of Physics,
Kansas State University, 116 Cardwell Hall, Manhattan, KS 66506,
USA}

\author{Arthur Kosowsky}
\email{kosowsky@pitt.edu} \affiliation{Department of Physics and
Astronomy, University of Pittsburgh, 3941 O'Hara Street,
Pittsburgh, PA 15260, USA}

\begin{abstract} A primordial magnetic field in the early universe
will cause Faraday rotation of the linear polarization of the
cosmic microwave background generated via Compton scattering at
the surface of last scattering. This rotation induces a non-zero
parity-odd (B-mode) polarization component. The Wilkinson
Microwave Anisotropy Probe (WMAP) 5-year data puts an upper limit
on the magnitude of the B-polarization power spectrum; assuming
that the B-polarization signal is totally due to the Faraday
rotation effect, the upper limits on the comoving amplitude of a
primordial stochastic magnetic field range from $6 \times
10^{-8}$  to $2 \times 10^{-6}$ G on a comoving length scale of 1
Mpc, depending on the power spectrum of the magnetic field.
\end{abstract}

\maketitle
\section{Introduction}

The cosmic microwave background temperature and polarization
anisotropy measurements presented
in the 5-year data of the Wilkinson Microwave Anisotropy Probe (WMAP) provide the
most comprehensive characterization to date of the perturbations present in the early
universe at the time of photon-baryon decoupling \cite{WMAP_5_meas}.
The microwave background polarization field is conventionally decomposed into a sum of
parity-even (``E-type'') and parity-odd (``B-type'') parts; since B-type
polarization is absent in a cosmological scenario
involving only scalar density perturbations,
it serves as a powerful test of other cosmological physics,
including primordial tensor and vector metric perturbations \cite{EBdecomp}.
Another source of B-type polarization is Faraday rotation of the
orientation of linear polarization due to the presence
of a cosmological magnetic field: if a purely E-type polarization pattern is everywhere rotated
by $\pi/4$, it becomes a purely B-type polarization. Smaller rotation angles generate
a component of B-type polarization from an initial pure E-type polarization field.

Faraday rotation of polarized radiation is in general a powerful
tool to estimate magnetic field amplitudes and spectra (see
\cite{Vallee} for a recent review). In the cosmological context,
Kosowsky and Loeb \cite{kosowsky96} proposed microwave background
Faraday rotation to probe a homogeneous primordial magnetic
field; this issue has been re-addressed by several authors for the
homogeneous field \cite{giovannini,Demianski:2007fz}, as well as
for a stochastic field \cite{dolgov,kklr05,giovannini05,GK08}. For
a homogeneous field, the same polarization rotation effect
additionally induces non-zero parity-odd cross correlations
between microwave temperature and B-polarization anisotropies and
E-polarization and B-polarizations anisotropies \cite{ferreira}
and correlations between different multipoles \cite{DKY98} which
are absent for standard cosmological scenarios. Primordial
magnetic fields are interesting because they could serve as seed
fields for the observed magnetic fields in galaxies and galaxy
clusters \cite{Widrow}.

Similar effects
(microwave background birefringence and parity-odd anisotropy spectra) may also
have non-magnetic explanations. In particular, microwave background
polarization plane rotation occurs in models with violation of parity
symmetry and/or Lorentz invariance in the early universe
\cite{reh}. Parity-odd polarization anisotropy spectra appear due
to gravitational lensing \cite{seljak} or
non-zero primordial magnetic helicity \cite{pogosian}.

Here we use WMAP 5-year B-polarization power spectrum upper limits
\cite{WMAP_5_spectra} to place limits from Faraday rotation
on a stochastic primordial magnetic field.
We assume that the cosmological magnetic field was
generated during or prior to the early radiation-dominated epoch.
The high conductivity of the primordial plasma  results in a
``frozen-in'' magnetic field, fixing the temporal
dependence to be the simple scaling ${\mathbf B}({\mathbf x},
\eta)={\mathbf B}({\mathbf x})/a^2$ where $a$ is the scale factor
and $\eta$ conformal time. Throughout this paper, ${\bf B}$
represents the comoving value of the magnetic field. We also
normalize the scale factor $a$ by setting $a_0 = 1$ today, and we
employ cosmological units with $\hbar = 1 = c$ and gaussian units
for electromagnetic quantities.

\section{CMB Faraday Rotation Effect}

Electromagnetic waves  propagating in a magnetized medium
have the plane of their linear polarization rotated (see, e.g.,
\cite{krall}).  A linearly polarized wave can be expressed as a
superposition of left  and right circularly polarized
waves. The magnetic field induces a phase velocity
difference between the two circular polarization, resulting in rotation of the
polarization plane. The rotation angle $\alpha$ for a plane wave with comoving
frequency $\nu_0$ propagating in the direction ${\bf n}$ satisfies
\begin{equation}
\frac{d\alpha}{d\eta} = \frac{3}{(4\pi)^2\nu_0^2 q}\dot\tau({\bf
x})~{\bf n}\cdot{\bf B}({\bf x})
\end{equation}
where $\dot\tau=x_e n_e\sigma_T a$ is the differential optical depth, $n_e$ and $x_e$
are the comoving electron number density and ionization fraction, $\sigma_T$ is the
 Thomson cross section, and $q$ is the magnitude of the electron charge. Faraday
rotation of the microwave background is a subtle problem, because the polarization is
generated and rotated simultaneously in the region of the last
scattering surface. However, as long as the total rotation is small
compared to $\pi/2$, the total rotated polarization angle can be
expressed simply as an average of the rotated polarization angle
from each infinitesimal piece of path length through the surface
of last scattering, neglecting depolarization effects \cite{kklr05}.

When considering a stochastic magnetic field, we make the simplifying approximation that any
magnetic field component with a wavelength shorter than the
thickness of the surface of last scattering is neglected: for
these components, the rotation of polarization generated at
different optical depths will tend to cancel, leaving little net
rotation. Then we can treat the magnetic field as constant
throughout the rotation region, so that the total rotation, which
is the sum over the rotations of each infinitesimal piece of
generated polarization, can be expressed as the total rotation
incurred by the polarization generated at some particular
effective optical depth (for details see \cite{kklr05}.)

A Gaussian random magnetic field is described by the two-point
correlation function in wavenumber space as
\begin{equation}
   \langle B^*_i({\mathbf k})B_j({\mathbf k'})\rangle
   =(2\pi)^3 \delta^{(3)}\!({\mathbf k}-{\mathbf k'}) P_{ij} P_B(k),
   \label{spectrum}
\end{equation}
where $P_B(k)$ is the magnetic field power spectrum,  vanishing
for all wavenumber larger than the damping scale $k_D$,
$P_{ij}\equiv\delta_{ij}-\hat{k}_i\hat{k}_j$ is the plane
projector, and  $\hat{k}_i=k_i/k$. We use the Fourier transform convention
$B_j({\mathbf k}) = \int d^3\!x \,
   e^{i{\mathbf k}\cdot {\mathbf x}} B_j({\mathbf x})$.
(We neglect
nonzero magnetic helicity since it does not affect the polarization
plane rotation \cite{ensslin03,dolgov,kklr05}.)
The power spectrum $P_B(k)$ is related to the total energy density
$E_B$ of the magnetic field through $E_B=\int_0^{k_D} d^3\! k
P_B(k) $,  and it is given by simple power law $P_B(k)=A_B
k^{n_B}$. Smoothing the field on a comoving scale $\lambda >
\lambda_D=2\pi/k_D$ by convolving with a Gaussian smoothing
kernel, the
smoothed magnetic field amplitude $B_\lambda$ is \cite{kklr05}
\begin{equation}
B_\lambda^2 \lambda^{n_B+3}=\frac{A_B~\Gamma\left(n_B/2 +
3/2\right)}{2\pi^2},
  \qquad\qquad \lambda>\lambda_D.
   \label{energy-spectrum-S}
\end{equation}
We assume that the magnetic field cut-off scale is determined by the
Alfven wave damping scale \cite{jedamzik98},
\begin{equation}
  \left({k_D \over {\rm Mpc}^{-1}}\right)^{n_B + 5}
  \approx 2\times 10^4
  \left({B_\lambda\over 10^{-9}\,{\rm G}}\right)^{-2}
  \left({k_\lambda\over {\rm Mpc}^{-1}}\right)^{n_B + 3},
  \label{damping-scale}
\end{equation}
which  will always be a much smaller scale than the Silk damping
scale (thickness of the last scattering surface) for standard
cosmological models.

For a given comoving radiation frequency $\nu_0$, consider the rotation angle
$\alpha({\bf n})$ of the microwave background linear polarization as a function
of sky direction ${\bf n}$.
For a stochastic magnetic field, it is obvious that the rotation angle averaged over the sky is zero.
Expanding the rotation angle in spherical harmonics
leads to the definition of the angular power spectrum $C_l^{\alpha}$ via
\begin{equation}
   \left\langle \alpha({\bf n}) \alpha({\bf n^\prime})\right\rangle
   = \sum_{l} \frac{2l+1}{4\pi} C_l^{\alpha} P_l ({\bf n \cdot n^\prime}).
\end{equation}
where $P_l(x)$ are Legendre polynomials. The rotation angle power spectrum
due to a stochastic magnetic field is \cite{kklr05}
\begin{equation}
   C_l^{\alpha} \simeq \frac{9 l(l+1)}{(4\pi)^3q^2\nu_0^4}
   \frac{B^2_\lambda}{\Gamma\left(n_B/2 +3/2 \right)}
   \left(\frac{\lambda}{\eta_0}\right)^{n_B+3}
   \int_0^{x_D}
   dx \, x^{n_B}j^2_l(x),
   \label{ClRR-sym-int}
\end{equation}
where $x_D=k_D\eta_0$, $j_l(x)$ are spherical Bessel functions.
This rotation multipole expression contains a sharp
short-wavelength cutoff $k_D$ in the magnetic field; in reality, the effective cutoff
will be smoothly spread over a range of scales.
Given this rotation angle power spectrum, we need to compute
the B-polarization power spectrum induced by the rotation field
from the primordial E-polarization.
This is given by Eq. (44) of Ref.~\cite{kklr05},
\begin{equation}
   C^{BB}_l = N_l^2 \sum_{l_1l_2}{(2l_1+1)(2l_2+1)\over 4\pi(2l+1)}
   N_{l_2}^2 K(l,l_1,l_2)^2 C^{EE}_{l_2}C^{\alpha}_{l_1}\left(C^{l0}_{l_10l_20}\right)^2
   \label{answer}
\end{equation}
where $C^{l0}_{l_10l_20}$ are Clebsch-Gordan coefficients
(with techniques for numerical evaluation in
Appendix B of Ref.~\cite{kklr05}), the normalization factor $N_l =
(2(l-2)!/(l+2)!)^{1/2}$ and the function $K(l,l_1,l_2)\equiv
-{1\over 2}\left(L^2 + L_1^2 + L_2^2 -2L_1L_2
-2L_1L +2L_1-2L_2 -2L\right)$ with $L\equiv l(l+1)$, $L_1\equiv l_1(l_1+1)$,
and $L_2\equiv l_2(l_2+1)$.
The cross-correlations between temperature and B-polarization and between
E and B-polarization are zero for a stochastic magnetic field with zero helicity.
The modifications to the existing E-polarization power spectrum and the cross-correlation
between E-polarization and temperature are negligible for small rotation angles, and we
ignore this second-order effect in our analysis.

\begin{figure}
\includegraphics[width=8.6cm]{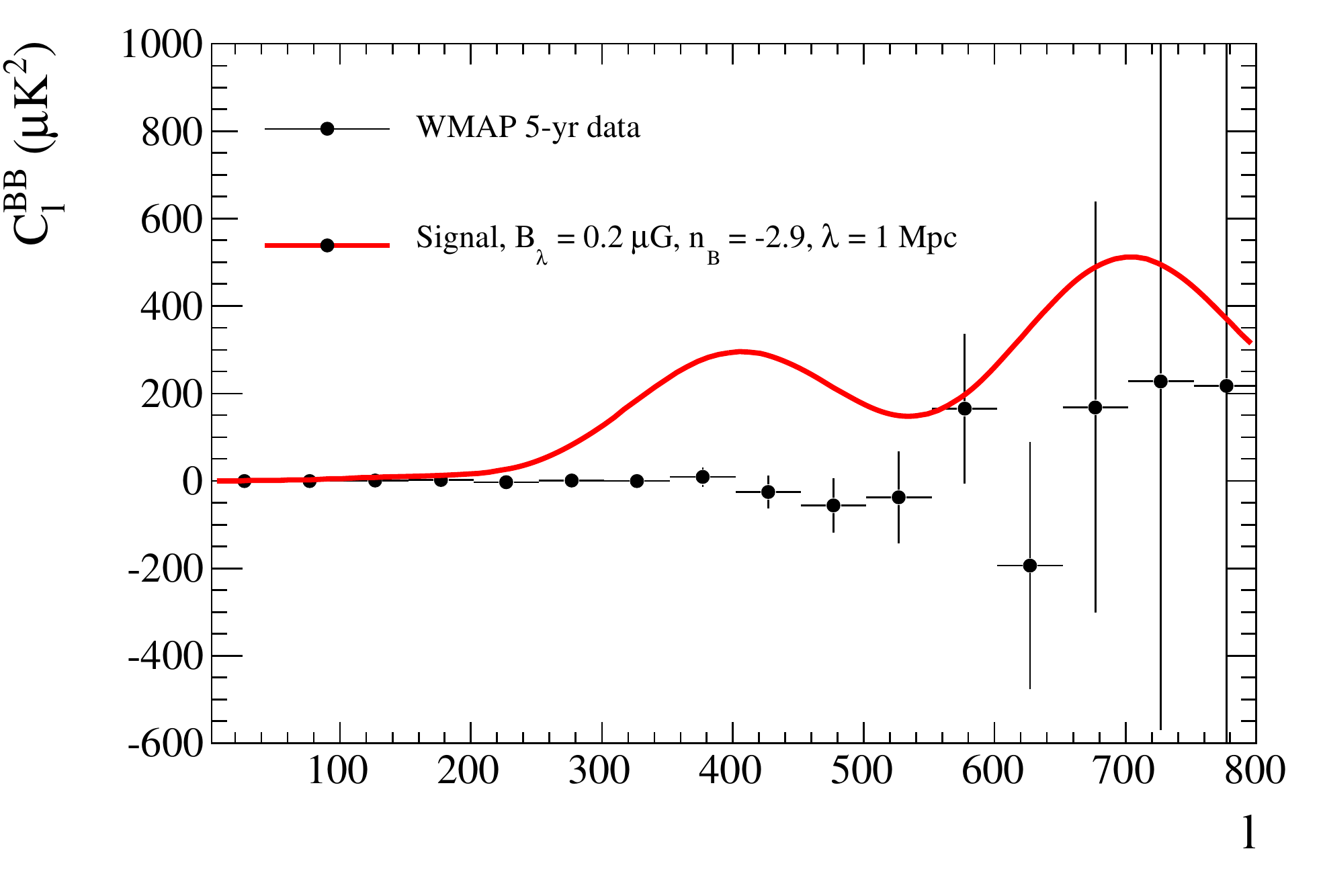}
\caption{The solid line shows the B-polarization power spectrum
due to Faraday rotation from the WMAP best-fit cosmology, plus a
stochastic magnetic field with amplitude $B_\lambda = 0.2$~$\mu$G,
$\lambda =1$~Mpc, $\nu = 30$ GHz, and power law index $n_B =
-2.9$. The data points show binned WMAP 5-year B-polarization
data with $l<800$. Note that this magnetic field amplitude and
power spectrum is ruled out by the data in the region between
$l=300$ and $l=500$. Points with $l<150$ do not contribute
significantly to the constraint. \label{fig:comparison_nb29_5nG}}
\end{figure}

\section{Results and Discussion}

To compare the data with the theory prediction, we calculate
$C_{l}^{BB}$ for $B_{\lambda} = 1$ nG, $\lambda = 1$ Mpc, $n_B$
between -2.9 and -1 in steps of 0.005, and $l$ values from 2 to
1000. Our range of spectral indices is based on plausible
magnetic field generation mechanisms \cite{ratra} and spectral
indices which are not excluded by other bounds; see
Refs.~\cite{n-limits,grasso}. The input $C_{l}^{EE}$ values up to
$l=4000$ were calculated using CMBFAST \cite{cmb} for the
best-fit WMAP cosmological model. The obtained set of
$C_{l}^{BB}$ is then further rescaled as $B_\lambda^2$ and
$\lambda^{n_{B}+3}$ to obtain theoretical $C_{l}^{BB}$ for
different values of magnetic field amplitude and scale,
respectively. WMAP data \cite{WMAP2} covers five frequency bands
with central frequencies ranging from 23 GHz to 94 GHz. We use
the measured B-polarization power spectrum limits for the three
highest-frequency WMAP bands centered at 41, 61, and 94 GHz (Q, V, and W);
our limits could be improved by including information from the
lowest two frequency channels, but these are dominated by
synchrotron foregrounds. We use only data for $l>32$, for which
we can treat measurement errors for different $l$ values as
uncorrelated. Most of the constraining power
of the data comes from multipoles between $l=200$ and $l=500$, as displayed
in Fig.~1, where
the theoretical signal begins to be comparable to the
errors on the data points; neglecting the data
with $l<32$ only marginally affects the limits obtained here.

\begin{figure}
\includegraphics[width=8.6cm]{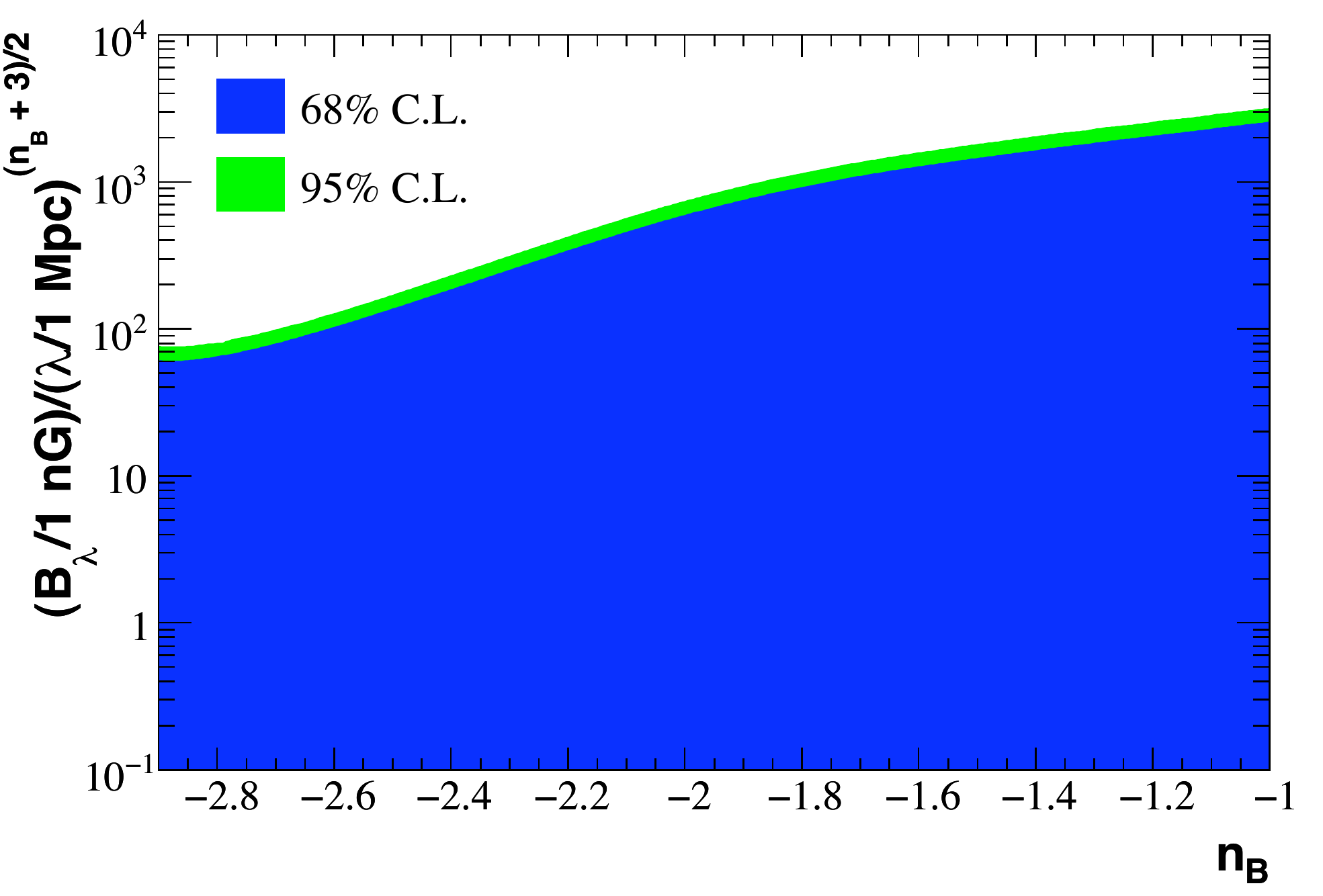}
\caption{The 68\% and 95\% C.L. limit bands on the $B_\lambda$ as
function of $n_B$ for $\lambda$ = 1 Mpc.}
\label{fig:bLimit_cl}
\end{figure}

\begin{figure}
\includegraphics[width=8.6cm]{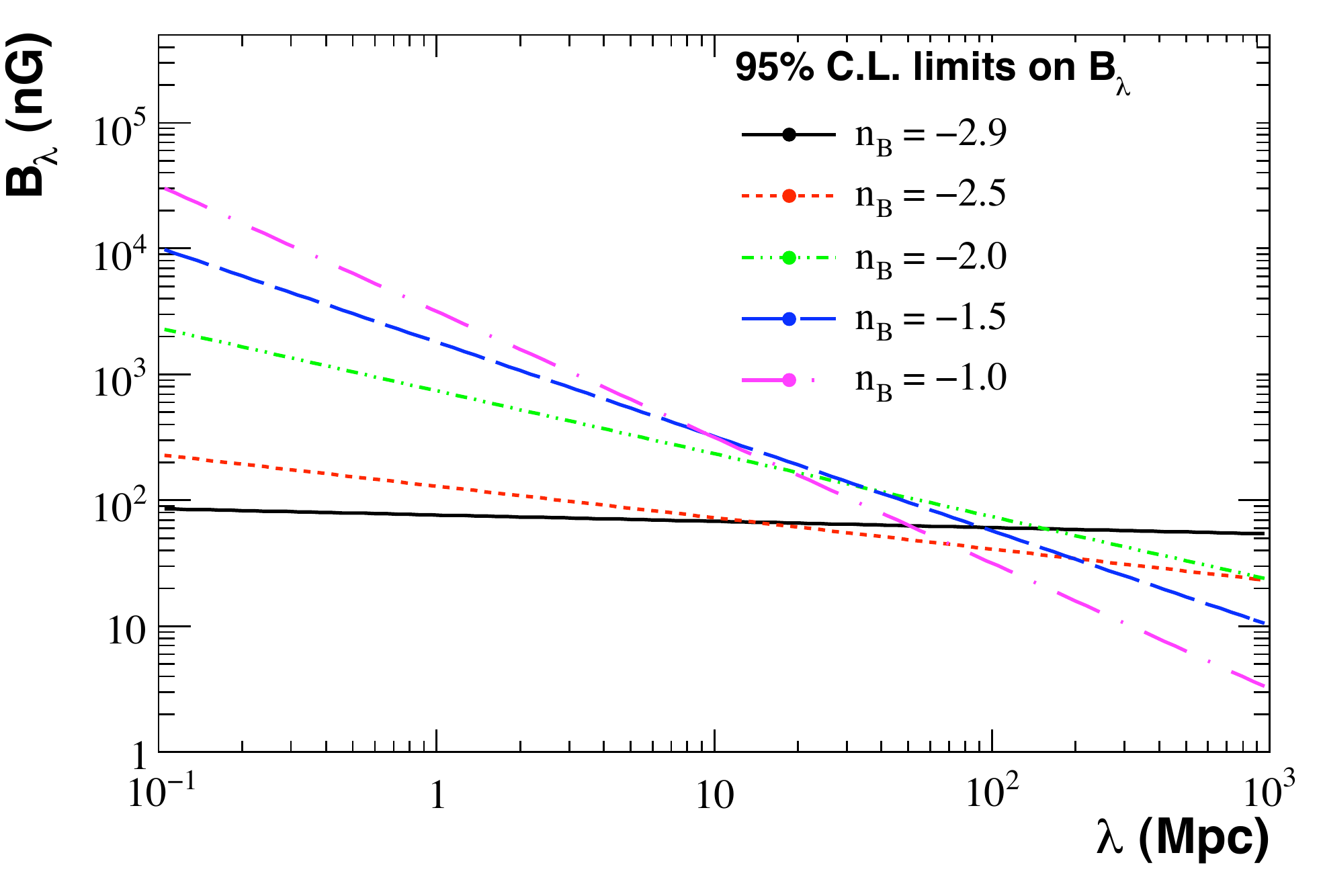}
\caption{95\% C.L.\ upper limits on $B_\lambda$ for fixed
values of $n_B=-2.9$, -2.5, -2.0, -1.5, and -1.0 as functions of
$\lambda$.} \label{fig:bLimit_vs_lambda}
\end{figure}

The theoretical  $C_l^{BB}$ power spectra are compared with the
WMAP data using $\chi^2$ statistics. We conservatively include an
extra 40\% error on theoretical $C_l^{BB}$ values to compensate
for using the nominal frequency center of each WMAP band rather
than the detailed frequency response of each band. Theoretical
uncertainties on calculated values of $C_{l}^{EE}$ are neglected,
since these are much smaller than the experimental uncertainties
on WMAP data. As the data is consistent with the hypothesis of
$C_l^{BB}=0$ (see Fig. \ref{fig:comparison_nb29_5nG}), we set
limits on values of $B_\lambda$ as a function of $n_B$ and
$\lambda$. The 68\% and 95\% confidence limit bands on $B_\lambda$
as a function of $n_B$ are given in Fig.~\ref{fig:bLimit_cl}. We
also display in Fig.~\ref{fig:bLimit_vs_lambda} the 95\%
confidence upper limits on $B_\lambda$ for fixed values of $n_B$
ranging from -2.9 to
 -1.0 as functions of $\lambda$.

These results presume that B-polarization signal is totally due
to the Faraday rotation effect. Other possible source of
B-polarization, such as inflationary gravitational
waves \cite{EBdecomp}, gravitational lensing \cite{seljak} or
non-zero primordial magnetic helicity \cite{pogosian}, will strengthen
the magnetic field limits derived here.

A primordial magnetic field on a scale $\lambda=100$ Mpc must have
an amplitude $B_\lambda$ of less than $10^{-7}$ G for any power
spectrum between $n_B=-2.9$ and $n_B=-1$. These limits are weaker
that those obtained recently  in Refs.\ \cite{GK08}. Improvements
on this limit using the Faraday rotation signal will be
challenging, requiring substantially more sensitive polarization
measurements at frequencies of 50 GHz or below; most current or
planned high-sensitivity polarization experiments use bolometer
detection technology, which tend to lose sensitivity in this
frequency range. Upcoming measurements from the LFI instrument on
Planck will be of roughly comparable polarization sensitivity to
the WMAP measurements at similar frequencies. One notable effort
is the QUIET experiment \cite{samtleben07}, which uses coherent
detector technology at frequencies of 40 and 90 GHz. It is
eventually anticipated to measure the BB polarization power
spectrum with an improvement in sensitivity over WMAP by a factor
of $10^4$.  The lowest frequency of 40 GHz, compared to 23 GHz
for WMAP, makes the Faraday polarization power spectrum smaller
by a factor of 7.3 for the same magnetic field; such an experiment
could thus place limits on the amplitude of a primordial field
which are a factor of approximately $(10^4/7.3)^{1/2}=37$ more
stringent than those here. A detailed analysis for this
experiment \cite{ferreira08} projects limits on a homogeneous
primordial field below $10^{-10}$ G; however, primordial field
limits will always be stronger than the stochastic field
considered here, because Faraday rotation of a primordial field
also generates nonzero TB and EB polarization cross-correlations
which are larger than the corresponding BB power spectrum.
Polarized foreground emission is also a potentially difficult
systematic limit to these measurements. Other upcoming
polarization experiments aimed at detecting the BB polarization
from inflation are likely to improve on the Faraday rotation
limits here, though the Faraday rotation power spectrum amplitude
decreases by a factor of  230 between WMAP's 23 GHz channel and a
likely 90 GHz lowest-frequency bolometer channel.

The cosmological magnetic field itself generates microwave
background anisotropies, particularly via the tensor and vector
perturbations it induces \cite{mkk} (see Refs.~\cite{grasso} for
overviews of magnetized cosmological perturbations). Here we have
neglected these perturbations, limiting ourself to considering the
Faraday rotation effect alone, which provides a distinctive
signature of magnetic fields nearly independent of other
properties of the universe. We use the direct observational data
without any priors.  Our magnetic field limits are weaker than
those arising from  CMB temperature maps \cite{lewis04,GK08} or
current bounds on a homogeneous magnetic field due to
correlations between different $l$ modes \cite{chen} and CMB
temperature non-gaussianity \cite{ss09}. The strongest future
limits on magnetic fields in the early universe will likely come
from limits on the temperature and polarization fluctuations
generated by magnetic field-induced vector perturbations
\cite{mkk}; we will address these current limits from WMAP data
elsewhere.

\acknowledgments
We thank Eiichiro Komatsu, Mike Nolta, and the WMAP Science
Collaboration for providing the WMAP power spectra for individual
frequency bands. Useful
discussions from Y. Gershtein, A. Gruzinov, G. Lavrelashvili, B.
Ratra, and L. Samushia contributed to this paper. We particularly appreciate
an anonymous referee who caught an incorrect normalization in an
earlier version of this work. T.K.\ acknowledges partial support from
INTAS 061000017-9258 and Georgian NSF grants ST06/4-096 and ST08/4-442,
and also from the
International Center for Theoretical Physics associate membership
program. T.K.\ and Y.M.\ acknowledge financial support from DOE
grant DE-FG02-99ER41093. A.K.\ has been partly supported by NSF
grant AST-0546035.

\end{document}